\documentclass[%
 reprint,
superscriptaddress,
 amsmath,amssymb,
 aps,
pra,
]{revtex4-2}

\usepackage{graphicx}
\usepackage{dcolumn}
\usepackage{bm}
\usepackage{upgreek}
\usepackage{siunitx}
\usepackage{float}
\usepackage{tikz}
\usepackage{amsmath}
\usepackage{xcolor}

\begin{document}

\preprint{APS/123-QED}

\title{High sensitivity accelerometry with a feedback-cooled magnetically levitated microsphere}

\author{Charles W. Lewandowski}
    \affiliation{Department of Physics, Montana State University, Bozeman, Montana 59717, USA}
\author{Tyler D. Knowles}%
    \affiliation{Department of Mathematics, West Virginia University, Morgantown, West Virginia 26506, USA}
\author{Zachariah B. Etienne}
    \affiliation{Department of Physics and Astronomy, West Virginia University, Morgantown, West Virginia 26506, USA}
    \affiliation{Center for Gravitational Waves and Cosmology, West Virginia University, Chestnut Ridge Research Building, Morgantown, West Virginia 26505, USA}
\author{Brian D'Urso}
    \email{durso@montana.edu}
    \affiliation{Department of Physics, Montana State University, Bozeman, Montana 59717, USA}
\date{\today}

\begin{abstract}

We show that a magnetically levitated microsphere in high vacuum can be used as an accelerometer by comparing its response to that of a commercially available geophone. This system shows great promise for ultrahigh acceleration sensitivities without the need for large masses or cryogenics. With feedback cooling, the transient decay time is reduced and the center-of-mass motion is cooled to \SI{9}{K} or less. Remarkably, the levitated particle accelerometer has a sensitivity down to \SI{3.6e-8}{\mathit{g}/\sqrt{Hz}} and gives measurements similar to those of the commercial geophone at frequencies up to \SI{14}{Hz} despite a test mass that is four billion times smaller. With no free parameters in the calibration, the responses of the accelerometers match within \num{3}\% at \SI{5}{Hz}. The system reaches this sensitivity due to a relatively large particle mass of \SI{0.25}{\upmu g}, a low center of mass oscillation frequency of \SI{1.75}{Hz}, and a novel image analysis method that can measure the displacement with an uncertainty of \SI{1.6}{nm} in a single image.
\end{abstract}

\maketitle

\section{Introduction}

High sensitivity accelerometry has myriad applications in fundamental and practical fields of physics and engineering. The ability to measure extremely small accelerations and forces has uses in absolute gravimeters~\cite{niebauer1995new, bidel2013compact, liu2017nanogravity}, inertial navigation~\cite{battelier2016development}, tests of quantum gravity~\cite{kafri2014classical, albrecht2014testing}, gravitational wave detection~\cite{abbott2017gw170817}, precision measurements of the Newtonian constant of gravitation~\cite{cavendish1798xxi} and other tests of fundamental physics~\cite{moore2018tests}.

Typical accelerometers are based on clamped resonator systems~\cite{gerberding2015optomechanical, bao2016optomechanical, guzman2014high}. With cryogenic temperatures, force sensitivities as low as $S_F^{1/2} \sim \SI{e-21}{N/\sqrt{Hz}}$ are predicted~\cite{moser2014nanotube}. Using a Si$_3$N$_4$ membrane~\cite{norte2016mechanical}, quality factors of $10^8$ can be achieved at room temperature with oscillation frequencies of $\sim \SI{150}{kHz}$, and thermal noise limited force sensitivities of $S_F^{1/2} \sim \SI{e-17}{N/\sqrt{Hz}}$ are possible. Mechanical devices have the advantage of typically being extremely compact~\cite{krause2012high, li2018characterization}. Systems with very test large masses, such as LISA Pathfinder, can have acceleration sensitivities of $S_a^{1/2} \sim \SI{e-16}{\mathit{g}/\sqrt{Hz}}$~\cite{armano2016sub} where $g$ is standard gravity, $g = \SI{9.8}{m/s^2}$.  Cold atom interferometry systems have also been proposed for measuring small changes in gravity~\cite{yu2006development, stern2009light, biedermann2008gravity} with acceleration sensitivities as low as $S_a^{1/2} \sim \SI{e-9}{\mathit{g}/\sqrt{Hz}}$~\cite{zhou2012performance, biedermann2015testing}.

Levitated systems avoid dissipation associated with the mechanical contact of the resonator with its environment. Force sensitivities of $S_F^{1/2} \sim \SI{e-16}{N/\sqrt{Hz}}$ and $S_F^{1/2} \sim \SI{e-18}{N/\sqrt{Hz}}$ have been measured with particles in optical traps~\cite{ranjit2015attonewton, ranjit2016zeptonewton}. Acceleration sensitivities of $S_a^{1/2} \sim \SI{e-10}{\mathit{g}/\sqrt{Hz}}$~\cite{timberlake2019acceleration} have been reported using a permanent magnet levitated above a superconductor at cryogenic temperatures.

Levitated optomechanical systems in vacuum provide extreme isolation from the environment, making them powerful candidates for high sensitivity accelerometry. The field has been dominated by optical trapping since its development by Ashkin and Dziedzic~\cite{ashkin1971optical}, in which feedback cooling is typically required for the levitated particle to remain trapped at pressures less than approximately $\SI{0.08}{Torr}$~\cite{monteiro2017optical, rider2018single}. A magnetic trap that does not rely on gravity for confinement has been demonstrated down to a pressure of $\sim \SI{0.1}{Torr}$~\cite{o2019magneto}. Magneto-gravitational traps have been developed~\cite{houlton2018axisymmetric, hsu2016cooling} and have exhibited stable trapping to a pressure of $\sim \SI{e-10}{Torr}$ with a feedback cooled center-of-mass motion from room temperature to $\SI{140}{\upmu K}$~\cite{klahold2019precision}. Recent cooling experiments in an optical trap have demonstrated a center-of-mass motion temperature of $\SI{50}{\upmu K}$ for large particles ($\approx \SI{10}{\upmu m}$)~\cite{monteiro2020force}. Cooling to the quantum ground state of a sub-micrometer particle has also been shown, reaching a temperature of $\SI{12}{\upmu K}$ from room temperature~\cite{delic2020cooling}.

\begin{figure*}
    \includegraphics[width=2.0\columnwidth]{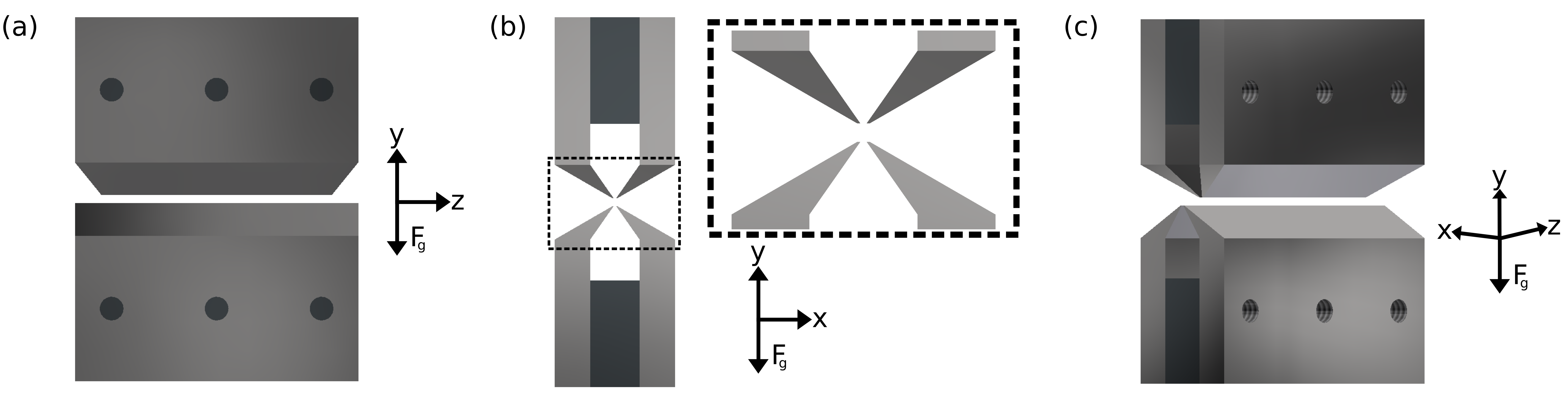}
    \caption{(a) The linear magneto-gravitational trap as viewed from the transverse ($x$) direction. The length of the bottom pole pieces is approximately $\SI{26}{mm}$. The top pole pieces are cut shorter to a length of approximately $\SI{21}{mm}$. The asymmetry combined with the force of gravity constrains the particle in the axial ($z$) direction. (b) The view of the trap as viewed from the axial direction showing the quadrupole symmetry in the transverse and vertical ($y$) directions. (c) A rotated view of the trap showing the quadrupole symmetry and the broken symmetry in the vertical direction. The tapped holes pictured on the pole pieces are used to attach the trap to its mount.}
    \label{fig:trap}
\end{figure*}

In this paper, we demonstrate levitation of a diamagnetic borosilicate microsphere in a magneto-gravitational trap down to a pressure of $\sim \SI{e-7}{Torr}$ at room temperature. The relatively large mass of the $\SI{60}{\upmu m}$ microsphere and low oscillation frequencies compared to optical trapping systems~\cite{lewandowski2019comparison} make this a promising optomechanical system for high sensitivity room temperature accelerometry. The center-of-mass motion is cooled with feedback to damp transients on a reasonable timescale. To check the calibration, accelerations are directly applied to the system via a surface transducer. 

A critical component of the system is a new offline image analysis technique we have developed to determine the displacement of the trapped particle from photos recording its motion over time.  In particular, we mitigate image background noise and avoid issues with fractional pixel translations by constructing a pixel-independent ``eigenframe'', against which we compute the cross correlation.

\section{Experimental Setup}

\subsection{Loading and Trapping of Microspheres}

The magneto-gravitational trap, designed with two samarium-cobalt (SmCo) permanent magnets and four iron-cobalt alloy (Hiperco-50A) pole pieces (see Fig.~\ref{fig:trap}), creates a three-dimensional potential well to stably trap diamagnetic particles.  The total potential energy of an object with volume $V$ of diamagnetic material with magnetic susceptibility $\chi$ and mass $m$ in an external magnetic field subject to standard gravity $g$ is 
\begin{equation}
    U = -\frac{\chi B^2 V}{2 \upmu_0} + mgy,
    \label{eq:potential}
\end{equation}
where $B = |\vec{B}|$ is the magnitude of the magnetic field, $\upmu_0$ is the vacuum permeability, and $y$ is the vertical displacement of the material~\cite{simon2000diamagnetic}. For diamagnetic materials ($\chi < 0$), a stable trap is formed at a magnetic field minimum in the absence of gravity.

The four pole pieces are configured in a quadrupole arrangement surrounding the two permanent magnets. The quadrupole field lies in the transverse-vertical ($x-y$) plane. Symmetry is broken in the vertical-axial ($y-z$) plane by cutting the top pole pieces shorter along the axial direction. This asymmetry along with gravity forms the trapping potential in the axial ($z$) direction~\cite{hsu2016cooling, slezak2018cooling}.

To reduce the effect of thermal noise while maintaining sensitivity to acceleration, larger trapped particles are preferred. A loading method has been developed to allow reliable trapping of large microspheres~\cite{klahold2019precision}. In these experiments, we chose borosilicate microspheres (Cospheric BSGMS-2.2 53-63um-10g) with greater than $90\%$ of particles in the diameter range of $\SI{53}{\upmu m}$-$\SI{63}{\upmu m}$. Insulating polyimide tape is attached to the tip of an ultrasonic horn~\cite{perron1967design} to electrostatically hold large microspheres to the tip. The ultrasonic horn shakes the particles off and into the trapping region at atmospheric pressure. An AC voltage is applied to two pole pieces while the other two are kept isolated from the AC voltage to form a linear quadrupole ion (Paul) trap~\cite{paul1990electromagnetic, douglas2005linear} for the particles that have non-zero net charge.

Note that Eq.~\ref{eq:potential} requires a large gradient in $B^2$ to balance gravity. The large dimensions of the magneto-gravitational trap form an extremely weak potential. The Paul trap is much stronger, allowing particles to be successfully levitated near the center of the trap. A DC voltage, typically between $\SI{20}{V}$ and $\SI{40}{V}$, is applied from the top to the bottom pole pieces to help counter gravity and center large particles in the trap.

\begin{figure}
    \includegraphics[width=.7\columnwidth]{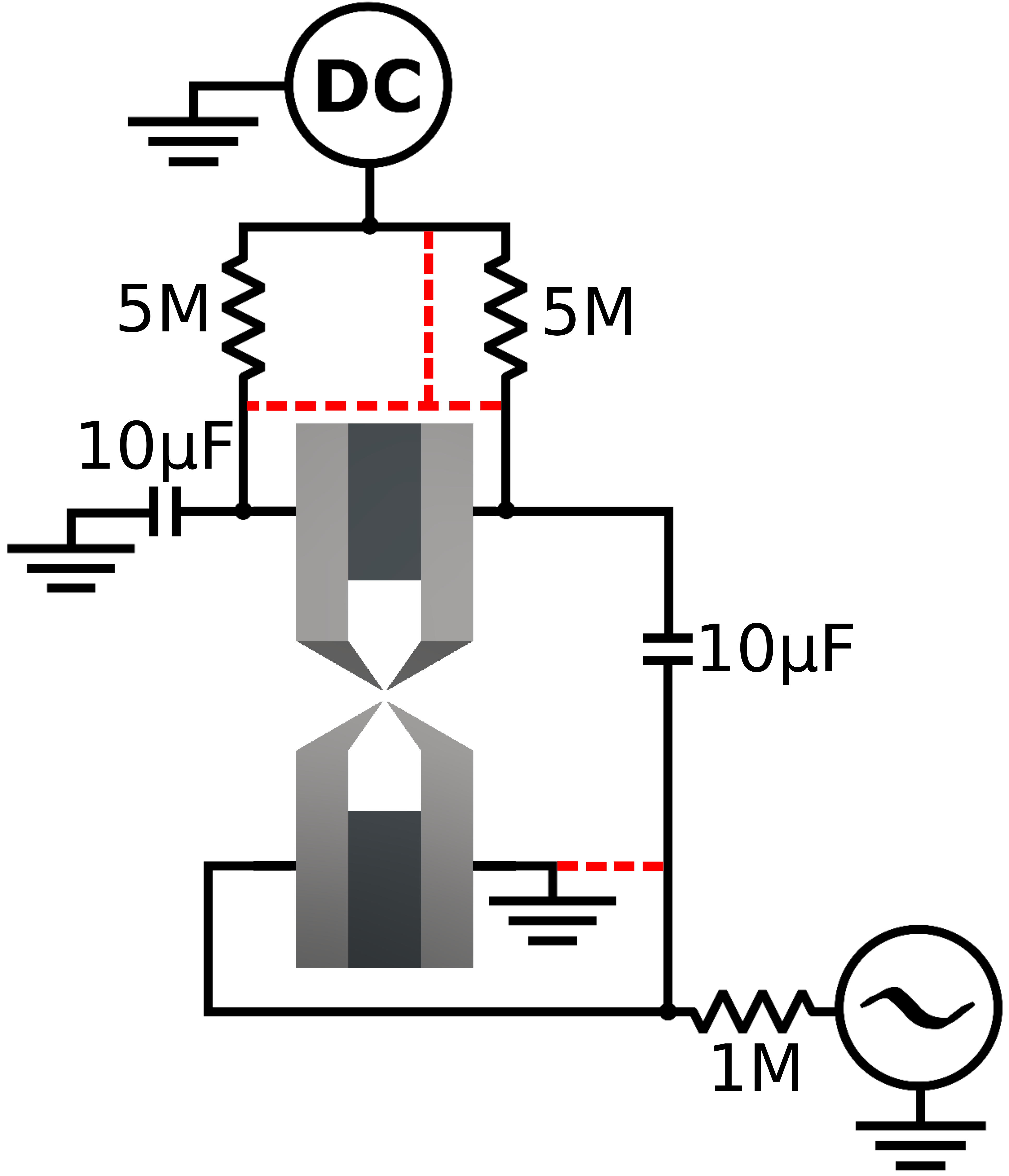}
    \caption{A simple schematic of the circuit used to apply the AC voltage for the Paul trap and the DC bias to help counteract gravity. A $\SI{50}{s}$ filter prevents the addition of high frequency noise. The red dashed lines indicate where jumpers are added to prevent image charge currents from going through high resistance paths when the Paul trap is not in use.}
    \label{fig:paul_trap}
\end{figure}

The DC voltage across the top and bottom pole pieces is supplied from a 1-ppm digital-to-analog converter (DAC, Analog Devices AD5791). The DAC is floated to a voltage between $\SI{-300}{V}$ and $\SI{0}{V}$ using a modified stacking of Texas Instruments REF5010 high-voltage references~\cite{instruments2013stacking} in steps of $\SI{5}{V}$. The voltage reference circuit can be modified to allow for positive voltages as well. The DAC allows for fine tuning of the voltage, and the resulting potential is estimated to be stable to $<\SI{3}{ppm}$.

After the particle is loaded into the hybrid Paul-magneto-gravitational trap, the Paul trap is turned off before pumping down the system to high vacuum. The AC voltage is slowly decreased while adjusting the DC voltage to keep the particle centered vertically in the magneto-gravitational trap. When the Paul trap is completely off, jumpers, indicated by the dashed lines in Fig.~\ref{fig:paul_trap}, are added to eliminate all of the high resistance paths for the movement of image charges.

A mechanical roughing pump along with a turbomolecular pump achieve a pressure of $\SI{e-7}{Torr}$ in the vacuum chamber. To eliminate vibrations from these pumps, they are closed off from the chamber and turned off while pumping continues with an ion-sputter pump. A pressure of $\sim \SI{e-7}{Torr}$ was maintained for all measurements reported.

\subsection{Table Stabilization}

Changes in the tilt of the optical table cause the equilibrium position of the levitated particle to shift. In the weak direction of the trap, very small changes in tilt can have a significant effect on the equilibrium position. For small tilts, the shift in equilibrium position is described by
\begin{equation}
    \Delta z_{eq} \approx \frac{g}{\omega_0^2}\Delta \theta.
    \label{eq:eq_shift}
\end{equation}

To avoid any large shifts in the equilibrium position, a method has been developed to feedback stabilize the relative tilt of the optical table in real time. The tilt of the table is measured with an ultra-high sensitivity tiltmeter (Jewell Instruments A603-C) and read on a computer. Using two mass flow controllers, air is added or removed from one side of the floating table to keep it level. 

Without stabilization, the relative tilt of the table can change by $\pm \SI{150}{\upmu rad}$ or more. For a levitated particle with an axial oscillation frequency $\omega_0/(2\pi) = \SI{2}{Hz}$, this corresponds to a $\pm \SI{9.5}{\upmu m}$ shift in equilibrium, which is much larger than the typical oscillation amplitudes due to environmental vibrations. With feedback stabilization, this value can be 200 times smaller, resulting in only negligible shifts in equilibrium position.

\subsection{Real-Time Image Analysis and Feedback Cooling}

\begin{figure}
    \includegraphics[width=1.0\columnwidth]{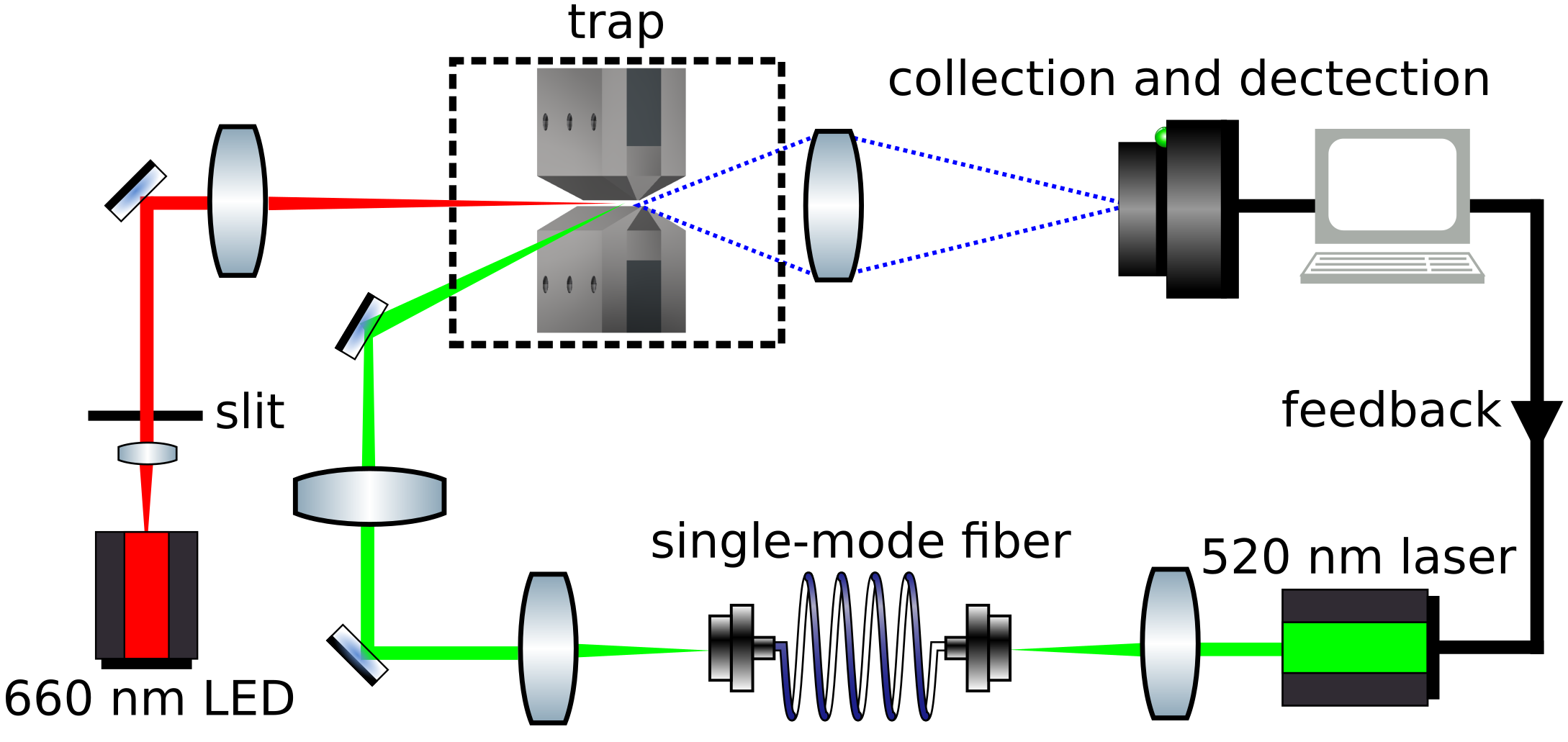}
    \caption{ 
    Light from a pulsed $\SI{660}{nm}$ LED illuminates a slit which is imaged onto the particle, as indicated by the red path. The control laser, indicated by the green path, utilizes a $\SI{520}{nm}$ diode laser. Radiation pressure from the laser applies a force that heats or cools the center-of-mass motion of the particle, depending on the phase of the drive relative to the motion of the particle. The scattered green light is blocked by a long-pass filter, while the scattered illumination light is collected and imaged onto a CMOS camera. The images are analyzed in real time to apply the feedback drive to the particle.}
    \label{fig:optics}
\end{figure}

The particle is stroboscopically illuminated using a $\SI{660}{nm}$ LED with a repetition rate of $\SI{100}{Hz}$ and a pulse duration of $\SI{1}{ms}$. As shown in Fig.~\ref{fig:optics}, light from the LED is collimated using an aspheric lens and passed through a $\SI{100}{\upmu m}$ slit. The slit is imaged onto the particle and magnified to illuminate the entire region of interest. The particle is imaged onto the CMOS camera with a 0.09 NA telecentric objective (Mitutoyo 375-037-1). All recorded images are 256 by 128 pixels, corresponding to a field of view of $\SI{300}{\upmu m}$ by $\SI{150}{\upmu m}$.

\begin{figure*}
    \includegraphics[width=2.0\columnwidth]{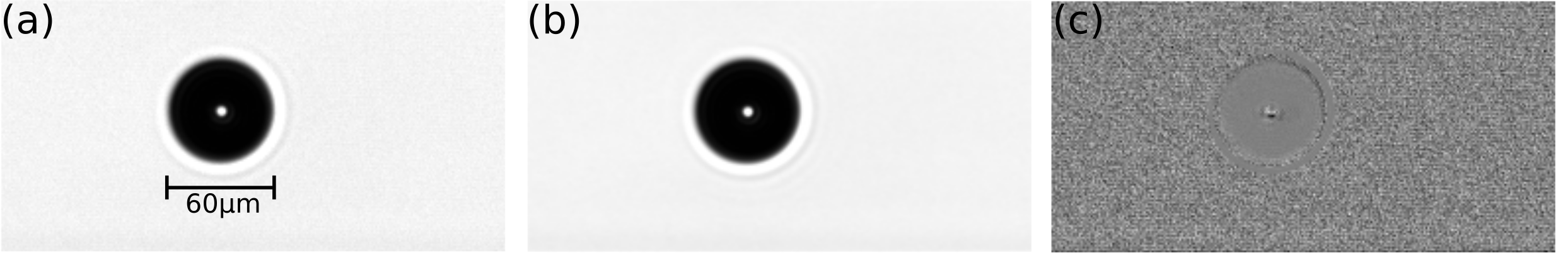}
    \caption{(a) Raw data: dark microsphere on light background with approximate scale. (b) Fifth eigenframe.  Note the smoothing of background features in comparison to the raw data. (c) Difference in pixel values between the zeroth eigenframe and the fifth eigenframe.  Note that grey pixels denote small differences; dark and light pixels represent the fifth eigenframe having a darker or brighter pixel than the zeroth eigenframe, respectively.  Differences in pixel values are scaled by a factor of 9.}
    \label{fig:Tyler_images}
\end{figure*}

As shown in Fig.~\ref{fig:Tyler_images}(a), the illuminated microsphere appears as a dark disk in each image (or frame). The microsphere diameter of approximately \SI{60}{\upmu m} corresponds to a diameter of approximately 60 pixels in each frame.  The microsphere never leaves the frame in the data we analyze.

The images from the CMOS camera are analyzed in real time to track the motion of the particle. Each image is thresholded to isolate the particle, and the apparent center-of-mass is calculated. The movement from frame to frame is used to calculate the velocity of the particle, which is then passed through a second order infinite impulse response (IIR) peak bandpass filter with a bandwidth of $\SI{1.0}{Hz}$ centered at $\SI{1.5}{Hz}$ to eliminate high frequency noise.

The measured and filtered velocity of the particle is used to damp and cool the center-of mass motion of the particle via algorithmic feedback~\cite{milatz1953reduction}. A damping force is applied to the particle using the radiation pressure of the light from a modulated $\SI{520}{nm}$ diode laser. Light from this control laser which scatters off the particle is blocked by a long-pass filter before the objective lens used for imaging. 

\section{Offline Image Analysis}

\begin{figure}
    \includegraphics[width=1.0\columnwidth]{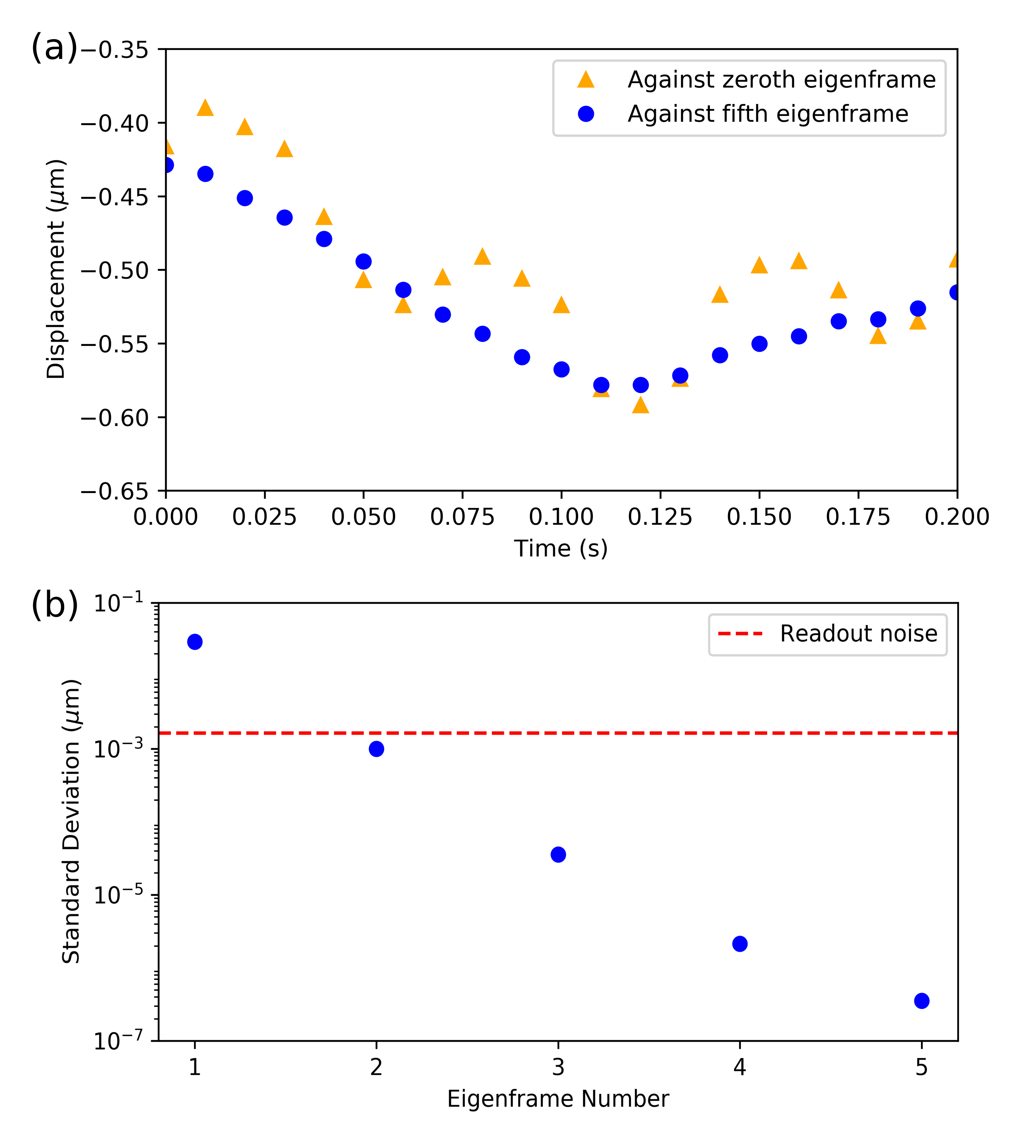}
    \caption{\label{fig:convergence_trajectory}(a) Displacement comparison between correlation against the zeroth and fifth eigenframes.  Note in particular the discontinuities appearing throughout the zeroth eigenframe displacement time series. (b) Standard deviation of displacement differences between eigenframes $n$ and $n-1$.}
\end{figure}

If limited to a resolution of one pixel, we could only track the microsphere's position to about 1 $\upmu$m. Sophisticated image analysis techniques exist, however, that measure displacement versus some reference frame to a small fraction of a pixel by incorporating \emph{all} pixel data from each frame. While the image analysis for feedback must be completed in real time, a more accurate but more computationally intensive algorithm can be used for offline analysis of the data. As our first approach, we adopted the cross-correlation function \texttt{register\_translation()} available in the \texttt{scikit-image} Python package~\cite{reg_trans_site,reg_trans_func} to determine the displacement of the particle relative to the first recorded frame of data (to which we refer as the ``zeroth eigenframe'').  While this approach largely seemed to work well, we noticed jump discontinuities in the microsphere displacement versus time as can be seen in Fig.~\ref{fig:convergence_trajectory}(a).

We attributed these discontinuities to noise in the zeroth eigenframe. As this frame was chosen arbitrarily, we anticipated that any other choice of reference frame would result in similar displacement discontinuities. To minimize the effects of this noise we devised a new ``eigenframe'' approach, which proceeds as follows: we first compute the translation in $z$ and $y$ of each frame against the zeroth eigenframe in the spatial domain using \texttt{register\_translation()}.  Using these translations, we line up all frames to their inferred displacement with respect to the zeroth eigenframe and construct a globally averaged frame.  We refer to the resulting averaged frame as the ``first eigenframe''. Specifically, the translations and averaging are performed using the two-dimensional discrete Fourier transforms of the images so that the choice of pixel alignment in the spatial domain does not result in loss of information.  The averaging smears out the noise present in the zeroth eigenframe and smooths the displacement data (as illustrated in Fig.~\ref{fig:convergence_trajectory}(a)). We then refine the translation values by correlating each frame against a translation of the first eigenframe (again in the Fourier domain) to the inferred particle location.  The resulting translations may be used to build a second eigenframe in a manner analogous to building the first, and this process can be iterated as many times as we like.

To further refine our position resolution, we modified \texttt{register\_translation()} to fit a slice of the correlation surface through the peak in the $z$-direction to a quadratic function using \texttt{SciPy}'s \texttt{optimize.curve\_fit()} function.  Locating the peak of this quadratic gives another estimate of the particle translation between each frame and the eigenframe.

To demonstrate that the translation values converge with eigenframe number, denote by $d_{n} \left( t_{i} \right)$ the axial displacement of the microsphere at time $t_{i}$ when correlated against eigenframe $n$ ($n=1,2,\hdots,5$).  We computed the standard deviation of $d_{n} \left( t_{i} \right) - d_{n-1} \left( t_{i} \right)$ over all $t_{i}$ (see Fig.~\ref{fig:convergence_trajectory}(b) and~\ref{fig:Tyler_images}(c)). Incredibly, the position differences quickly reach a standard deviation of less than \SI{1}{nm}, thus falling well below the physical resolution limit.  After repeating this eigenframe procedure five times, the standard deviation of the change in displacements drops to below \SI{1}{pm}. As this is far below other sources of displacement error in our experiment, the fifth eigenframe is the final one we compute.

\begin{figure}
    \includegraphics[width=1.0\columnwidth]{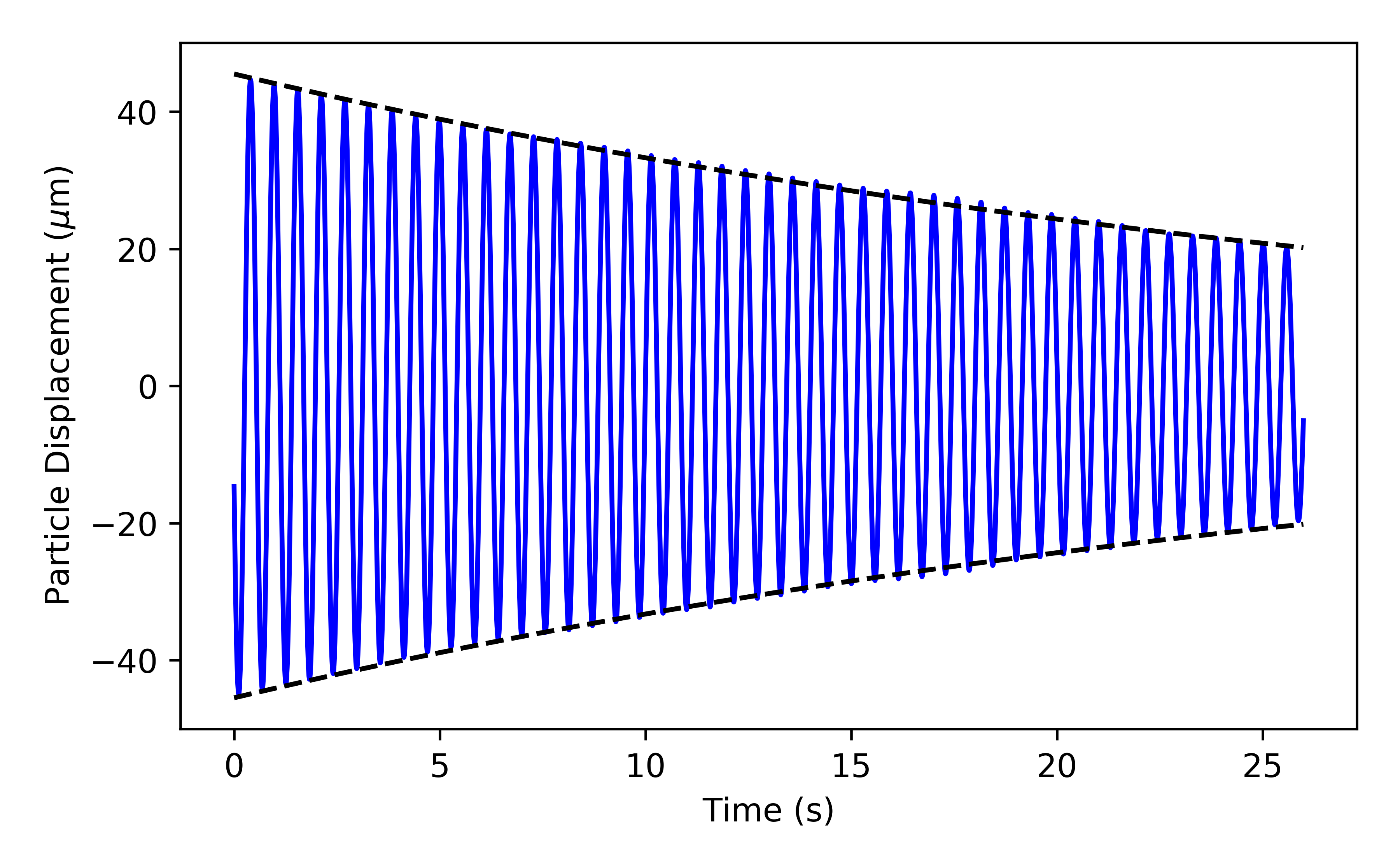}
    \caption{\label{fig:decay}Transient motion of the particle after excitation with feedback cooling applied. Analysis of the motion gives an axial ($z$) oscillation frequency of $\omega_0/(2\pi)=\SI{1.75}{Hz}$ and the decay rate is $\Gamma = \SI{6.26e-2}{s^{-1}}$ (black dashed lines). The amplitude range plotted and analyzed is chosen so that vibrational noise is negligible.}
\end{figure}

\section{Acceleration Measurement}

\begin{figure*}
    \includegraphics[width=2.0\columnwidth]{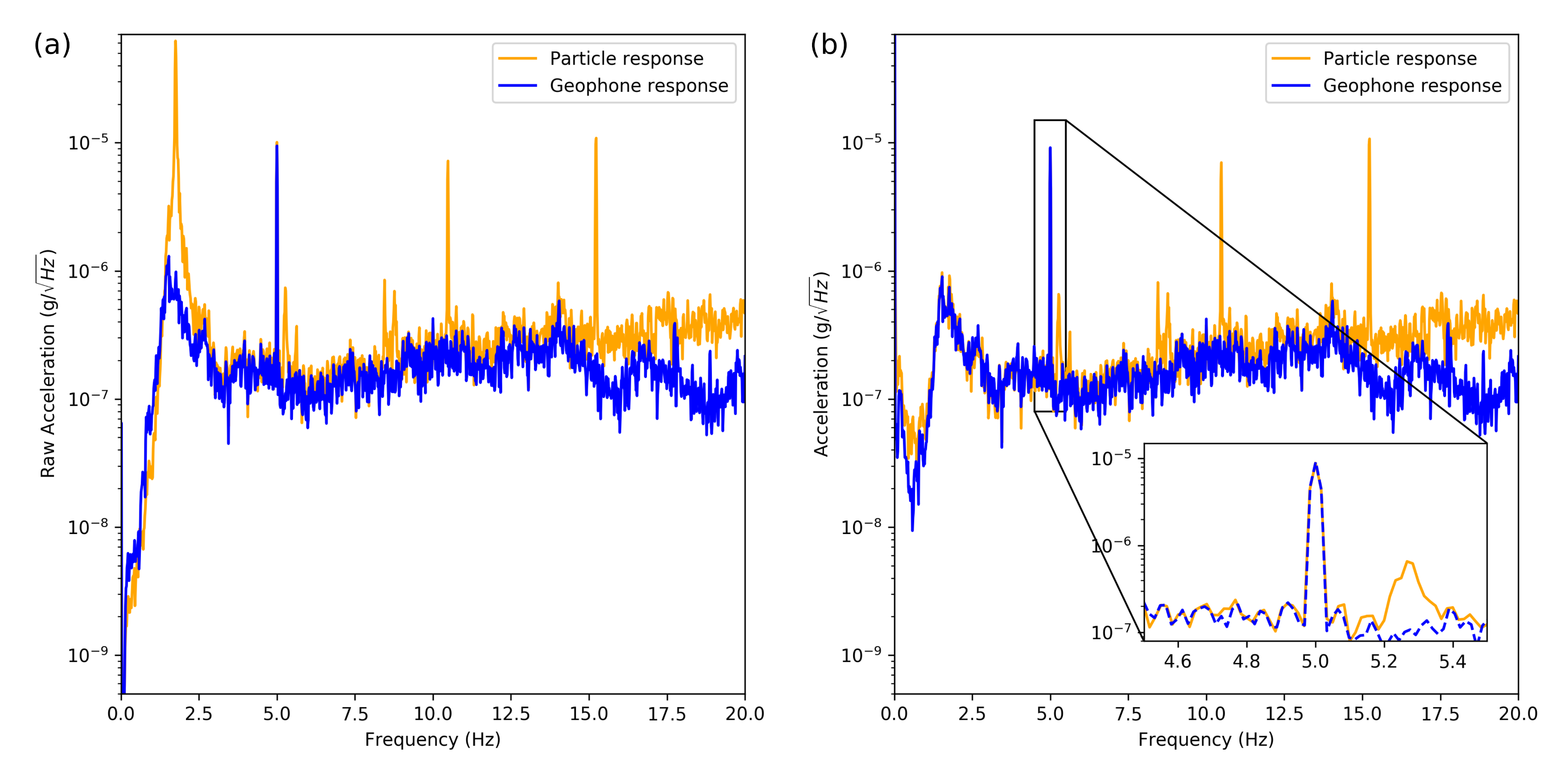}
    \caption{\label{fig:data_plot}(a) The raw data of the particle response in units of acceleration compared to the raw data of the L-4C geophone response in acceleration units with a $\SI{5}{Hz}$ drive to verify the calibration. The large peak at $\SI{1.75}{Hz}$ is the axial motion and the peaks at $\SI{10.5}{Hz}$ and $\SI{15.2}{Hz}$ are due to the transverse and vertical motions of the particle, respectively. (b) The particle and geophone responses from (a) divided by their harmonic oscillator responses. The amplitude of the peaks at $\SI{5}{Hz}$ differ from each other by less than 3\%. The blue line in the inset is dashed so both peaks are visible.}
\end{figure*}

We measure the acceleration sensitivity of the trapped particle by examining the effect of movement of the pneumatically isolated optical table (on which the trap and optics are mounted) on the particle. In the frame of the laboratory, consider the displacement of the particle in the axial direction, $z$, and the displacement of the camera, $z_0$. The camera directly measures $z^{\prime} = z - z_0$. The equation of motion for the particle in the laboratory frame is then
\begin{equation}
    \Ddot{z} + \Gamma\Dot{z} + \omega_0^2z = \Gamma\Dot{z_0} + \omega_0^2z_0
    \label{eq:eom}
\end{equation}
where $\Gamma$ is the damping rate and $\omega_0$ is the resonant angular frequency of the particle.

The displacement of the optical table, for example, from vibrations, can be written as an integral over all frequencies,
\begin{equation}
    z_0(t) = \int_{-\infty}^\infty d\omega^\prime A_0(\omega^\prime)\sin(\omega^\prime t + \phi)
    \label{eq:z_0}
\end{equation}
where $A_0$ is the strength of the drive as a function of frequency. 

After substituting Eq.~\ref{eq:z_0} into Eq.~\ref{eq:eom}, we can take the Fourier transform of Eq.~\ref{eq:eom}. Simplifying the resulting expression, we find that the magnitude of the transfer function is
\begin{equation}
    \left|\frac{Z^\prime(\omega)}{A_0(\omega)}\right| = \frac{\omega^2}{((\omega_0^2 - \omega^2)^2 + \Gamma^2\omega^2)^{1/2}}
    \label{eq:transfer_function}
\end{equation}
where $Z^{\prime}(\omega)$ is the Fourier transform of the particle's motion with respect to the camera. 

The minimum acceleration that can be detected for an oscillator in thermal equilibrium at temperature $T$ is~\cite{yasumura2000quality}
\begin{equation}
    S_a^{1/2} = \sqrt{\frac{4k_B\Gamma T}{m}}
    \label{eq:a_sense}
\end{equation}
where $k_B$ is Boltzmann's constant, $m$ is the mass, and $\Gamma$ is the damping rate of the oscillator. Feedback cooling at best keeps $\Gamma T$ constant, damping out potentially long-lived transients without a significant impact on sensitivity~\cite{geraci2010short}.

\subsection{Results}

A borosilicate microsphere was levitated with a DC bias across the vertical gap of the magneto-gravitational trap of $\SI{-37.2313}{V}$. Throughout the measurements, a vacuum pressure of $\SI{e-7}{Torr}$ was maintained and the tilt of the optical table was stabilized to within $\pm \SI{0.75}{\upmu rad}$. With the measured resonant frequency of the microsphere, Eq.~\ref{eq:eq_shift} gives that the equilibrium position of the particle was stabilized to within $\pm\SI{60}{nm}$.

Before acquiring acceleration data, the system magnification, a critical calibration parameter, is measured. By analyzing the recorded image of a USAF1951 calibration target (Edmund Optics \#58-198) through the system optics, the scaling factor $S_c = \SI{1.15}{\upmu m/pixel}$ was determined. For frequency calibration, the digital delay generator used to control all of the timing in the experiment is tied to a rubidium frequency standard (Stanford Research Systems, Inc.~FS725).

In order to eliminate any free parameters of the system, the transient response of the microsphere was measured after a small excitation in the axial direction, shown in Fig.~\ref{fig:decay}. The resonant frequency of the particle was measured to be $\omega_0/(2\pi) = \SI{1.75}{Hz}$. While feedback cooling the center-of-mass motion of the microsphere, the damping rate was measured to be $\Gamma = \SI{6.26e-2}{s^{-1}}$.

For comparison, we also place an L-4C geophone (Sercel, Inc.~\cite{sercel}) on the optical table. The sensitivity of this instrument and other critical parameters are given by the manufacturer. We added an additional amplification circuit with a gain of approximately 180 to boost the signal before digitization (modeled after that in~\cite{kirchhoff2017huddle}). 

The response of the particle to movement of the optical table is tested by applying $\SI{5}{Hz}$ sinusoidal drive with a surface transducer, oriented to push the table in the axial direction. While applying this external drive, a set of five $\SI{60}{s}$ measurements were recorded. Each measurement consists of \num{6000} images from the CMOS camera, which are analyzed with the algorithm described above. The averaged spectra of the resulting particle acceleration over five data sets is shown in Fig.~\ref{fig:data_plot}(a). For comparison, the measured acceleration of the test mass of the geophone is shown on the same plot. The vibration between $\SI{1}{Hz}$ and $\SI{2}{Hz}$ is believed to be a resonance of the optical table and overlaps with the resonance of the particle, causing an on-resonance excitation illustrated by the large peak at $\SI{1.75}{Hz}$. The transverse and vertical motion of the particle are at $\SI{10.2}{Hz}$ and $\SI{15.2}{Hz}$, respectively, likely creating peaks at the corresponding frequencies due to misalignments in the system.

To calculate the acceleration of the optical table from the acceleration of the particle and geophone test masses, the harmonic oscillator response of each is divided out of the raw data, resulting in the table acceleration shown in Fig.~\ref{fig:data_plot}(b). The two spectra match over a broad frequency range of approximately \SI{14}{Hz}. The amplitude of the two peaks at the external drive frequency, $\SI{5}{Hz}$, are within 3\% of each other, confirming the calibration between the two systems. Above $\SI{14}{Hz}$, the geophone response diverges from the particle response due to increasing noise in the particle acceleration measurement.

\subsection{Noise Analysis}

\begin{table*}
    \centering
    \begin{tabular}{|p{2cm}|p{8cm}|p{3cm}|p{3cm}|}
        \hline
        Parameter & Description & Geophone Value & Particle Value\\ 
        \hline\hline
        $T$ & Temperature & $\SI{300}{K}$ & $\SI{9}{K}$\\ 
        $\omega_0/2\pi$& Resonant frequency of oscillator & $\SI{.97}{Hz}$ & $\SI{1.75}{Hz}$\\
        $m$ & Mass of oscillator & $\SI{.957}{kg}$ & $\SI{2.5e-10}{kg}$\\
        $Q$ & Quality factor of oscillator & 1.845 & 175\\
        $S_g$, $S_c$ & Sensitivity (Note different units) & $\SI{281.7}{Vs/m}$ & $\SI{1.15}{\upmu m/pixel}$ \\
        $R_c$ & Resistance of geophone coil & $\SI{5546}{\Omega}$ &\\
        $S_g$ & Sensitivity of geophone oscillator & $\SI{281.7}{Vs/m}$ &\\
        $G_a$ & Gain of amplification circuit & 180.2 & \\
        $N_V$ & Input-referred voltage noise & $\SI{8.8}{nV/\sqrt{Hz}}$ &\\
        $N_A$ & Input-referred current noise & Negligible &\\
        $I_0$ & Energy density of scattered light & & $\SI{2.76e-4}{J/m^2}$\\
        $\Delta z$ & Readout noise & & $\SI{160}{pm/\sqrt{Hz}}$\\
        \hline
    \end{tabular}
    \caption{Critical parameters for the amplified L-4C geophone and levitated particle accelerometers. The geophone values are from the datasheets of the L-4C geophone and OPA188 operational amplifier used in the geophone amplifier.}
    \label{table:seisometer_noise_parameters}
\end{table*}

\begin{figure}
    \includegraphics[width=1.0\columnwidth]{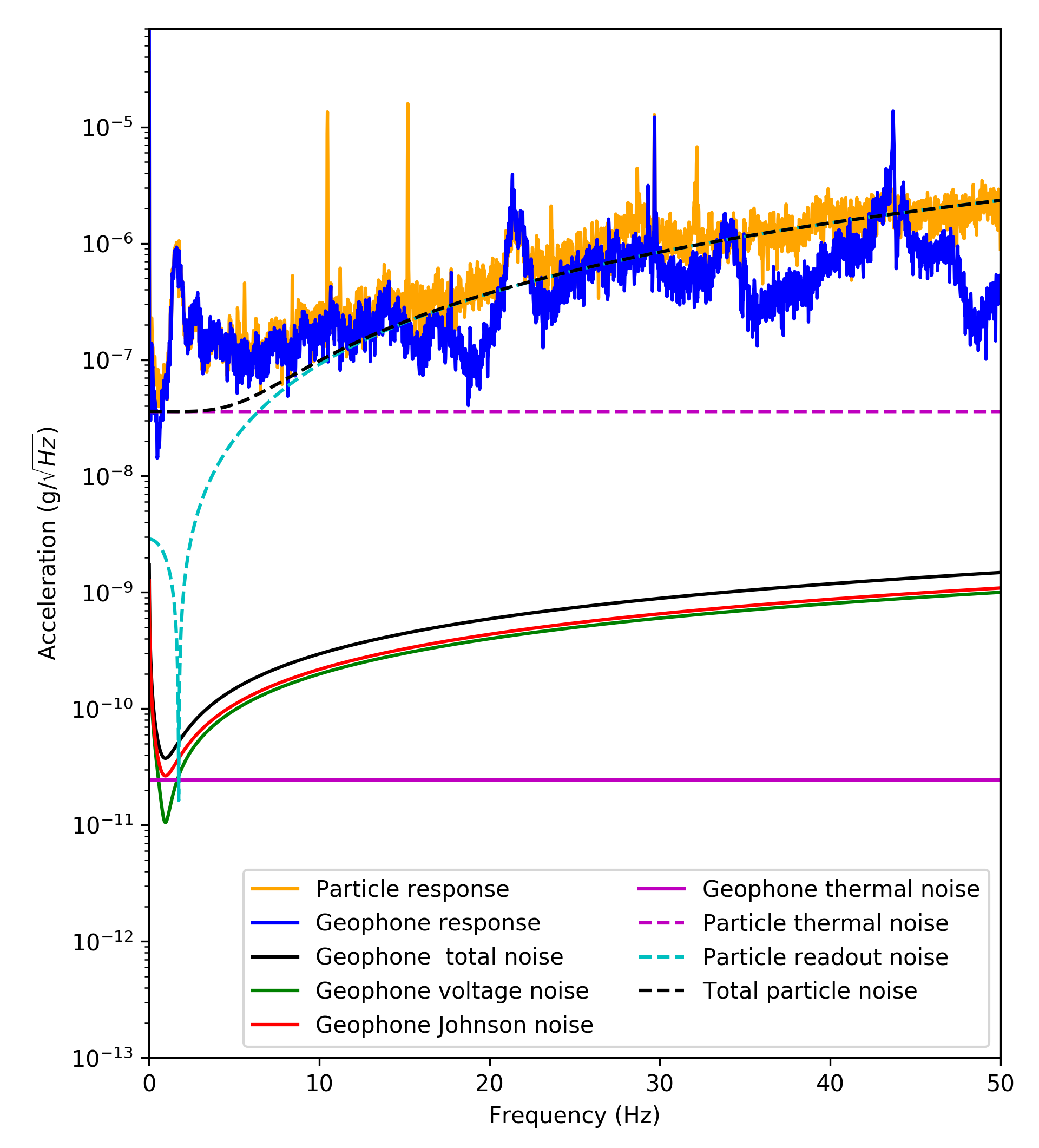}
    \caption{\label{fig:noise_0_Hz}The particle and geophone responses divided by the harmonic oscillator responses with no drive. Contributions to the noise for the geophone and particle are also shown.}
\end{figure}

The noise contributions for both the geophone and particle are plotted in Fig.~\ref{fig:noise_0_Hz} along with the (undriven) acceleration of the table as determined by the geophone and levitated particle.

The noise of the L-4C geophone and its accompanying amplification circuit can be broken down into four terms~\cite{kirchhoff2017huddle}. As displacement equivalent noise sources, they are:

\begin{equation}
    \textrm{n}_{\textrm{therm}}(\omega) = \sqrt{\frac{4k_B T\omega_0}{m Q}}\frac{1}{\omega^2}
    \label{eq:pend_noise}
\end{equation}

\begin{equation}
    \textrm{n}_{\textrm{Johnson}}(\omega) = \frac{\sqrt{4k_B T R(\omega)}}{G(\omega)}
    \label{eq:Johnson_noise}
\end{equation}

\begin{equation}
    \textrm{n}_{\textrm{voltage}}(\omega) = \frac{N_V(\omega)}{G(\omega)}
    \label{eq:volt_noise}
\end{equation}

\begin{equation}
    \textrm{n}_{\textrm{current}}(\omega) = \frac{N_A(\omega)R(\omega)}{G(\omega)}
    \label{eq:current_noise}
\end{equation}

The thermal noise of the damped harmonic oscillator is given by Eq.~\ref{eq:pend_noise} where $T$ is the temperature of the oscillator, $m$ is the mass, $\omega_0$ is the resonant angular frequency, and $Q$ is the quality factor. The thermal fluctuations are approximately $\SI{2.4e-11}{\mathit{g}/\sqrt{Hz}}$ for the geophone with the parameters listed in Table~\ref{table:seisometer_noise_parameters}. The Johnson noise of the geophone coil is given by Eq.~\ref{eq:Johnson_noise}, where $R$ is the real part of the complex impedance of the coil given by
\begin{equation}
    R(\omega) = R_c + \frac{iS_g^2\omega}{\omega_0^2 - \omega^2 + i\Gamma\omega}
    \label{eq:impedence}
\end{equation}
and where $R_c$ is the resistance of the coil, $S_g$ is the sensitivity of the oscillator, and $\Gamma$ is the damping rate of the oscillator. The harmonic oscillator response $G$ is given by 
\begin{equation}
    G(\omega) = \frac{\omega^3S_g}{\sqrt{(\omega_0^2 - \omega^2)^2 + \Gamma^2\omega^2}}.
    \label{eq:response}
\end{equation}
The input voltage and current noise of the amplification circuit is given by Eq.~\ref{eq:volt_noise} and Eq.~\ref{eq:current_noise}, respectively. $N_V$ is the input-referred voltage noise of the OPA188 operational amplifier~\cite{opa188} used in the amplification circuit, assumed to be constant over the frequency range of interest. The current noise of the amplification circuit is negligible compared to all other noise sources for the geophone. 
The noise sources add in quadrature to give the total noise of the geophone system $\textrm{n}_{\textrm{total}, g}$ as
\begin{equation}
    \textrm{n}_{\textrm{total}, g}^2 = \textrm{n}_{\textrm{therm}}^2 + \textrm{n}_{\textrm{Johnson}}^2 + \textrm{n}_{\textrm{voltage}}^2 + \textrm{n}_{\textrm{current}}^2.
\end{equation}

The noise of the levitated particle accelerometer has two contributions. First, the thermal noise of the particle is given by Eq.~\ref{eq:pend_noise}, where the parameters are now that of the particle (given in Table~\ref{table:seisometer_noise_parameters}). With feedback cooling applied, we measure the damping rate to determine the $Q$ of 175, but the effective temperature may be significantly reduced relative to the ambient temperature. Inspection of the minimum signal recorded (around \SI{.5}{Hz}) puts an upper limit on the noise of $\SI{3.6e-8}{\mathit{g}/\sqrt{Hz}}$ and a limit on the effective temperature associated with the damping of the particle of \SI{9}{K}.

The second noise source is readout noise from the camera and image analysis which is expected to be dominated by shot noise of the light and camera noise. To place a lower bound on the readout noise, we consider the precision to which a diffraction limited spot can be determined in the presence of shot noise.  This is described by
\begin{equation}
    \left< \left(\Delta z\right)^2 \right> = \frac{\sigma_{\textrm{PSD}}^2}{N}
\end{equation}
where $\sigma_{\textrm{PSD}}$ is the standard deviation of the point spread function (PSF) of the imaging optics and $N$ is the number of photons collected, or in our case, blocked, by the particle. 

For the lower bound on the noise, the PSF is calculated for a diffraction limited spot. The standard deviation is $\sigma_{\textrm{PSD}} = 0.225\lambda / \textrm{NA}$ where $\lambda$ is the wavelength of the scattered light and NA is the numerical aperture of the collection objective. For our system, $\sigma_{\textrm{PSD}} = \SI{1.65}{\upmu m}$. The number of photons is estimated from the brightness of the illumination in the CMOS camera and the number of pixels blocked by the \SI{30}{\upmu m} radius particle, resulting in an uncertainty in the location of the point of $\left<(\Delta z)^2\right> \approx \SI{0.4}{nm}$. Given that the particle is much larger than the diffraction limit, the readout noise is expected to be significantly higher than this.

From $\SI{14}{Hz}$ to $\SI{50}{Hz}$, the levitated particle acceleration spectrum is not vibration limited. Instead, it follows the expected shape of readout noise, which is a white noise source (in displacement) with the harmonic oscillator response divided out. We fit the spectrum in the frequency range of $\SI{35}{Hz}$ to $\SI{50}{Hz}$ to find the apparent readout noise of $\Delta z = \SI{1.6}{nm}$ per image or $\SI{160}{pm/\sqrt{Hz}}$, reasonably above the point source diffraction limit. The two noise sources add in quadrature, so that the total noise of the particle response is
\begin{equation}
    n_{\textrm{total}, p}^2 = n_{\textrm{therm}}^2 + \left(\Delta z\right)^2. 
\end{equation}

\section{Discussion}

We have experimentally demonstrated levitation of a $\SI{2.5e-9}{kg}$ borosilicate microsphere in high vacuum. This system shows great promise for ultrahigh acceleration sensitivities without the need for large masses or cryogenics. Feedback cooling reduces the transient decay time of the system, while also cooling the center-of-mass motion. With no free parameters in the calibration, the acceleration determined from the apparent motion of the particle both follows that of a commercial geophone below $\SI{14}{Hz}$ and matches the response to an external drive within 3\% at $\SI{5}{Hz}$, despite the particle having a mass that is $4 \times 10^{9}$ times smaller than the test mass in the geophone.

The sensitivity limit in the levitated particle accelerometer is estimated to be below $\SI{3.6e-8}{\mathit{g}/\sqrt{Hz}}$ at low frequencies, limited by either by the vibrations being measured or thermal noise associated with damping at \SI{9}{K}; a quieter environment would be needed to unambiguously determine the limiting factor and the effective temperature. Much lower center-of-mass temperatures have been reached with trapped particles in other systems, so there is room for significant improvement. For example, feedback cooling to $\SI{140}{\upmu K}$ in a magneto-gravitational trap~\cite{klahold2019precision} and $\SI{50}{\upmu K}$ in an optical trap~\cite{monteiro2020force} have been demonstrated. Lower center-of-mass temperatures in the current system could result in a sensitivity improvement of at least an order of magnitude, and might be reached by using a more precise real-time image analysis system for feedback cooling. Further improvements are possible using an even lower center-of-mass oscillation frequency or a higher camera frame rate. This high-sensitivity, self-calibrating system with negligible test mass may be particularly valuable for space-based accelerometry at low frequencies.

\begin{acknowledgments}
We acknowledge Lin Yi from JPL for discussions on potential mission requirements. This work was supported by the National Science Foundation under awards PHY-1707789, PHY-1757005, PHY-1707678, PHY-1806596, and OIA-1458952; the National Aeronautics and Space Administration under awards ISFM-80NSSC18K0538 and TCAN-80NSSC18K1488; and a block gift from the II-VI Foundation.  Offline data analysis was completed on the Spruce Knob Super Computing System at West Virginia University (WVU), which is funded in part by the National Science Foundation EPSCoR Research Infrastructure Improvement Cooperative Agreement \#1003907, the state of West Virginia (WVEPSCoR via the Higher Education Policy Commission), and WVU.
\end{acknowledgments}

\bibliography{apssamp}

\end{document}